\documentclass[fleqn,twoside]{article}
\usepackage{espcrc2}

\usepackage{graphicx}
\title{Radiative Upsilon Decay at the Endpoint}

\author{Adam K.~Leibovich\address{Theory Group, 
	Fermi National Accelerator Laboratory,\\ 
        Batavia, IL 60510, USA}%
        \thanks{I would like to thank my collaborators with whom this
		work was completed:
		C.~W.~Bauer, C.~W.~Chiang, S.~Fleming and I.~Low.
		I was supported in part by the Department of 
		Energy under Grant DE-AC02-76CH03000.}}
       
\begin{document}

\begin{abstract}
The standard NRQCD power counting breaks down and the OPE gives rise
to color-octet shape functions at the upper endpoint of the photon
energy spectrum in radiative $\Upsilon$ decay.  Also in this kinematic
regime, large Sudakov logarithms appear in the octet Wilson
coefficients, ruining the perturbative expansion.  Using SCET, the
octet shape functions arise naturally and the Sudakov logarithms
can be summed using the renormalization group equations.  We derive an
expression for the resummed octet energy spectrum.
\vspace{1pc}
\end{abstract}

\maketitle

%\section{Introduction}

The decay of the $\Upsilon$ in the endpoint region is interesting for
a number of reasons.  Firstly, it is a way to test Non-Relativistic QCD
(NRQCD) \cite{NRQCD} in decays, which could be important for
extractions of $\alpha_s$ \cite{gk}.  Secondly, in the endpoint
region, it is necessary to introduce another effective field theory,
Soft-Collinear Effective Theory (SCET) \cite{SCET}, due to the jet of
collinear particle that is produced in this kinematic region.
Finally, CLEO will obtain much more precise data on the first
three $\Upsilon$ resonances, so it is timely to see if we can predict
the decay spectrum.

The $\Upsilon(1S)$ decays radiatively about 3\% of the time.  When the
photon is emitted with maximum energy, $z\equiv 2 E_\gamma/M_\Upsilon
\to 1$, it is back-to-back with a jet of particles.  It is this
kinematic region that we are interested in predicting \cite{bcfll}.
Before discussing this region in particular, it will be useful to
review how the calculation is done in general.

To calculate the inclusive photon spectrum, it is advantageous to use
NRQCD.  The effective field theory (EFT) is based on the fact that the
$b$ quark is heavy, $m_b \gg \Lambda_{\rm QCD}$.  This implies that
the relative velocity of the the heavy quarks inside the bound state
is small, $v\ll1$.  For $\Upsilon$, $v^2\approx0.1.$ It is therefore
beneficial to do a double expansion in $v$ and $\alpha_s$.
Furthermore, it is possible to use NRQCD to prove the short and long
distance physics factorize in $\Upsilon$ decay \cite{NRQCD}.

The decay rate in NRQCD is written as
\begin{equation}
\frac{d\Gamma}{dz} = \sum_n C_n(z) \langle O(n)\rangle.
\label{untrunc}
\end{equation}
The $C_n$ are Wilson coefficients, calculable in perturbation theory,
while the $\langle O(n)\rangle$ are non-perturbative NRQCD matrix
elements (MEs).  The MEs schematically are
\begin{equation}
\langle O(n)\rangle = 
\langle \Upsilon|
 \psi^\dag\Gamma^{\prime n} \chi  \chi^\dag\Gamma^n \psi
|\Upsilon\rangle,
\end{equation}
where the $\Gamma^n$ can contain derivatives, color and spin
matrices, which can be classified in spectroscopic notation and as a
color-singlet or color-octet.  For instance $\langle
O_8(^1S_0)\rangle$ is a ME where $\Gamma^n$ contains only
the color-matrix $T^a$.

The series in Eq.~(\ref{untrunc}) is infinite, so to have any
predictive power, we need to truncate.  This is possible using the
velocity scalings of the MEs.  Each ME scales
as a certain power of $v$, depending on what terms from the NRQCD
Lagrangian needs to be inserted to have a non-zero overlap.  For the
endpoint spectrum, the relevant MEs are
\begin{eqnarray}
\langle O_1(^3S_1)\rangle &\sim& v^0,\\
\langle O_8(^1S_0)\rangle &\sim& v^4,\\
\langle O_8(^3P_0)\rangle &\sim& v^4.
\end{eqnarray}
The color-singlet ME $\langle O_1(^3S_1)\rangle$ can be related to the
wavefunction at the origin, and the decay rate through this channel is
the result obtained in the Color Singlet Model \cite{CSM}.

At first glance, it appears that the color-octet contributions to the
rate are tiny, $v^4\approx 0.01$, compared to the color-singlet.
However, that is not true.  To see why, we need to compare the rate
for each channel, including the Wilson coefficients.  For the
color-singlet rate, the $\Upsilon$ decays to a photon and two gluons,
so $C_1^{(0)}(^3S_1)\propto\alpha_s^2$.  For the color-octet channels,
on the other hand, the final decay products are a photon and one
gluon.  So $C_8^{(0)}(^1S_0)\sim C_8(^3P_0)\propto\pi\alpha_s$, where
the $\pi$ comes from there being one less particle in the final state.
The color-octet is enhanced perturbatively by a factor of $\pi\alpha_s
= {\cal O}(10)$.

But that is not all.  Since there are only two particles in the final
state for the color-octet decay, the rate is peaked at the endpoint,
$C_8^{(0)}(^1S_0)\sim C_8^{(0)}(^3P_0)\propto\delta(1-z)$.  If we
compare the integrated rate in the endpoint region, $(1-v^2 < z < 1)$,
we have $\Gamma^{\rm end}_1(^3S_1) \propto \alpha_s^2 v^2$ for the
color-singlet, and $\Gamma^{\rm end}_8(^1S_0)\sim\Gamma^{\rm
end}_8(^1S_0) \propto \pi\alpha_s v^4\approx \alpha_s^2 v^2$ for the
octet, using the fact that numerically $v^2\approx\alpha_s/\pi$.  So in
the endpoint region the singlet and octet rates contribute
equally.

If we went to higher order in the velocity expansion, we would find
further contributions, formally suppressed by higher powers of
$v$, that contribute equally in the endpoint region.  This is due to a
breakdown of the non-perturbative expansion.  The solution has been
known for some time \cite{RW}.  The series needs to be reordered as a
twist expansion, similar to what is done in $B$ decays \cite{shape}.  
The rate is then written as
\begin{equation}
\frac{d\Gamma}{dz} = \int dk_+ C_n(z,k_+) f_n(k_+) \langle O(n)\rangle,
\end{equation}
where $f_n(k_+)$ are shape functions, which measure the probability
for the quark pair (in state $n$) to have light-cone momentum
$k_+$.  As these are non-perturbative functions, they must be modeled.

We also must worry about the perturbative series.  The fact that the
color-octet rate begins as a $\delta$ function is already worrisome.
Higher order corrections could lead to further problems.  This in fact
is true.  The next-to-leading order (NLO) perturbative corrections
have been calculated for the color-singlet (numerically) \cite{kramer}
and color-octet \cite{MP}.  The NLO color-octet rate is singular at
the endpoint.  In particular, Sudakov logarithms appear in the Wilson
coefficients at NLO of the form
\begin{equation}
\alpha_s^2 \left(\frac{\log(1-z)}{1-z}\right)_+\quad
{\rm and}\quad \alpha_s^2 \left(\frac{1}{1-z}\right)_+,
\end{equation}
which become large as $z\to1$.  If we again integrate over the
endpoint, $(1-v^2 < z < 1)$, these terms give contributions to the
rate $\alpha_s^2 \log^2(v^2)$ and $\alpha_s^2 \log(v^2)$.  Both are
${\cal O}(\alpha_s)$, since $\alpha_s\log^2(v^2)$ and
$\alpha_s\log(v^2)$ are ${\cal O}(1)$.  This is an indication that the
perturbative series breaks down in the endpoint.  If we looked at
higher order perturbative corrections, we would get terms of the form
$\alpha_s[\alpha_s^n\log^{2n}(v^2)]$, which are also ${\cal
O}(\alpha_s)$.

The break down of the perturbative and non-perturbative series are
due to the same problem: NRQCD does not contain all of the
correct degrees of freedom for the endpoint region.  It is missing
collinear modes.  To correctly describe the physics at the endpoint we
need to couple NRQCD to an EFT that contains these missing fields,
SCET \cite{SCET}.

The invariant mass of the hadronic jet in the endpoint region, $M_X^2
= (1-z) M_\Upsilon^2 \equiv\lambda^2M_\Upsilon^2$, is much larger than
the energy in the jet, $(E_X = z M_\Upsilon/2)$.  We therefore have a
multiscale system, and EFT techniques are useful to
separate the scales.  In this case, the expansion parameter $\lambda =
\sqrt{1-z}$.  There are three classes of particles that are
included in the EFT, depending on the scaling of the momentum.  They
are massless quarks and gluons with:
\begin{eqnarray}
{\rm Collinear:} && E + p^3 \sim M_\Upsilon,
\ E - p^3 \sim \lambda^2 M_\Upsilon,\nonumber\\
&& p^{1,2} \sim \lambda M_\Upsilon,\nonumber\\
{\rm Soft:} && p^\mu \sim \lambda M_\Upsilon,\nonumber\\
{\rm Ultrasoft:} && p^\mu \sim \lambda^2 M_\Upsilon.\nonumber
\end{eqnarray}

The scales in the problem are the hard scale, $M_\Upsilon$, the
collinear scale, $\lambda M_\Upsilon$, and the soft scale, $\lambda^2
M_\Upsilon$.  The plan of attack is to integrate out the hard scale by
matching to the EFT and then run to the collinear scale using the
renormalization group equations (RGEs) of SCET.  At this point all
collinear particles are far off shell and should be integrated out.
This is done by matching onto the soft theory, which is same as the
large energy effective theory \cite{LEET}.  We will then run from the
collinear scale to the soft scale, at which point all the large
logarithms will be in the coefficient functions.  The logs in this
case are $\log\lambda\to\log(1-z)$, so by using the EFT RGEs, we have
resummed all the logs of $(1-z)$, i.e.\ the Sudakov logs \cite{bcfll}.

\begin{figure}[tb]
\includegraphics[width=7.5cm]{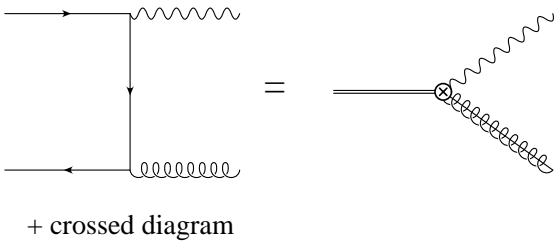}
\caption{Matching onto SCET.  The diagram on the left is calculated in
full QCD, and then expanded in $\lambda$ and $v$.  This is matched
onto an operator in the EFT.}
\label{fig:match}
\end{figure}

To follow this program, we first need to match onto SCET.  This is
done by calculating the graphs in QCD and expanding in $\lambda$ (and
in $v$ which will also match onto NRQCD).  An example of this is shown
in Fig.~\ref{fig:match}.  The matching, at leading order in $\lambda$,
gives two operators in the EFT, for the octet $^1S_0$ and $^3P_0$
channels.  We do not get any other operators at leading order, which
means there are no leading Sudakov logs in the color-singlet channel
\cite{hautmann}.  There are other operators suppressed by $\lambda$
\cite{BFL}.

\begin{figure}[tb]
\includegraphics[width=8.1cm]{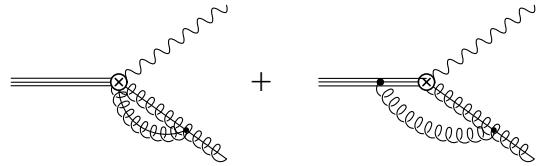}
\caption{The one-loop diagrams needed to calculate the anomalous
dimensions of the SCET operators.}
\label{fig:loop}
\end{figure}

We next need to run using the SCET RGEs.  This requires the anomalous
dimensions of the operators.  We therefore need to calculate the
one-loop diagrams in Fig.~\ref{fig:loop}.  The anomalous dimensions of
the $^1S_0$ and $^3P_0$ operators are the same, which is related to
the fact that the Sudakov logs are the same for both channels
\cite{MP}.  Please see \cite{bcfll} for the details.

\begin{figure}[b]
\includegraphics[width=7.8cm]{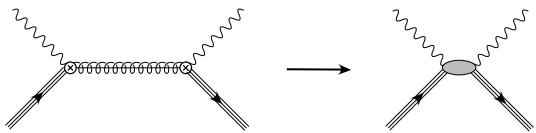}
\caption{Matching onto the soft theory, by doing an OPE.}
\label{fig:ope}
\end{figure}

We next match onto the soft theory, by removing all collinear
particles.  However, we have a collinear gluon in the final state, so
to remove it, we perform an operator product expansion (OPE).  The
result is a non-local operator, separated along the light-cone.  This
is shown diagrammatically in Fig.~\ref{fig:ope}.  
The operator we match onto is of the form
\begin{equation}
O_n(x) = \psi^\dag\Gamma^{\prime n} \chi 
         \delta(1-x+iD^+/M_\Upsilon) 
         \chi^\dag\Gamma^n \psi,
\end{equation}
where the derivative is in the $+$ light-like direction.  The rate is
now 
\begin{equation}
\frac{d\Gamma}{dz} = \int dx \sum_n C_n(x-z) f_n(x) \langle O(n)\rangle,
\end{equation}
where $f_n(x) = \langle O_n(x)\rangle/\langle O(n)\rangle$.  This is
just the rate including the shape function.  By using SCET, the shape
function appears naturally.

\begin{figure}[tb]
\includegraphics[width=7.4cm]{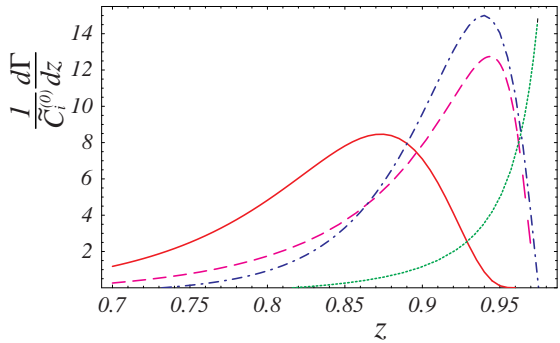}
\caption{The differential decay spectra near the endpoint region in
arbitrary units. The curves are described in the text.}
\label{fig:res1}
\end{figure}

We now run down to the soft scale.  For the details, see \cite{bcfll}.
At this point, there are no large logs in the operators.  We have
summed all the Sudakov logs into the Wilson coefficients.  To plot the
result, we use the model of \cite{KN}, introduced for $B$ decays.  For
$\Upsilon$, we need to give the shape function some unknown first
moment of order $\Lambda_{\rm QCD}$ \cite{bcfll}.  The results are
shown in Fig.~\ref{fig:res1}.
The dashed curve is the resummation without the shape function.  The
dotted curve is the singular terms in the one-loop result, and the
dot-dashed curve is these terms convoluted with the shape function.
The solid curve is the resummation convoluted with the shape function.
Note that the shape function without resummation and the
resummation without the shape function give similar results.  However,
both are necessary.

\begin{figure}[tb]
\includegraphics[width=7.4cm]{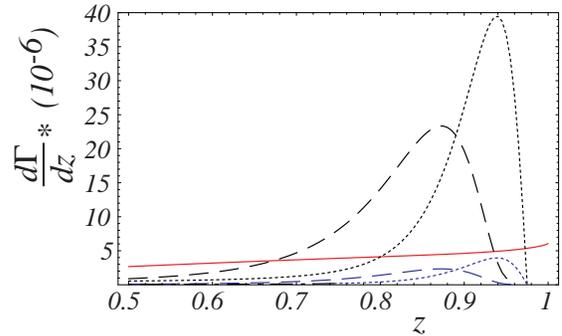}
\caption{The different channels for the spectra near the endpoint
region. The curves are described in the text.}
\label{fig:res2}
\end{figure}

At this point we compare to the color-singlet results.  To do this we
need the color-octet MEs.  In Fig.~\ref{fig:res2}, the solid curve is
the color-singlet result.  The dotted curve is the resummation without
the shape function, the dashed is resummation with the shape function.
The two sets of curves correspond to scaling the octet MEs from the
singlet by factors of $v^4/10$ and $v^4/100$.  As can be seen, to get
comparable results, the naive scaling ($v^4$) seems to be off by a
factor of 100 \cite{PCGMM}.

However, before a meaningful comparison to the data is possible, we
should follow a similar program for the color-singlet rate \cite{BFL}.
This will allow us to resum subleading Sudakov logs.  The existence of
these logs was first pointed out in \cite{hautmann}, where it was
observed that though the leading logs cancel in the color-singlet
differential rate, they are present in the derivative of the rate.

\end{document}